\documentclass[letterpaper,twoside,twocolumn,english,prl]{revtex4}
\usepackage[T1]{fontenc}
\usepackage[latin9]{inputenc}
\usepackage{amsmath,xcolor}
\usepackage{esint}
\usepackage{graphicx,amsfonts}
\usepackage{placeins}
\usepackage{float}
\usepackage{subfigure}
\usepackage{xcolor,hyperref}
\makeatletter

\@ifundefined{textcolor}{}
{%
 \definecolor{BLACK}{gray}{0}
 \definecolor{WHITE}{gray}{1}
 \definecolor{black}{rgb}{1,0,0}
 \definecolor{GREEN}{rgb}{0,1,0}
 \definecolor{black}{rgb}{0,0,1}
 \definecolor{CYAN}{cmyk}{1,0,0,0}
 \definecolor{MAGENTA}{cmyk}{0,1,0,0}
 \definecolor{YELLOW}{cmyk}{0,0,1,0}
}

\usepackage{babel}

\makeatother

\usepackage{babel}
\begin{document}

\title{Human wealth evolution is an  accelerating expansion underpinned by 
a decelerating optimization process}

\author{Paolo Sibani$^{1,\ast}$, Steen Rasmussen$^{1,2\ast}$ and Per Lyngs Hansen$^3$}
\affiliation{$^{1,\ast}$Center for Fundamental Living Technology (FLinT) \\
Department for Physics, Chemistry and Pharmacy, University of Southern Denmark\\ 
 $^{2}$Santa Fe Institute, Santa Fe, NM 87501, USA \\
 $^3$Department of Mechanical and Production Engineering, \\
 Aarhus University, Denmark\\$^\ast$ paolo.sibani@sdu.dk  and steen@sdu.dk}

\begin{abstract} \noindent  
{Optimization and expansion are two modes of  staged  evolution of complex systems 
where macroscopic observables change at a decreasing, respectively increasing, rate.
The number of microscopic variables and their interactions
are fixed in the first case but change dynamically in the second.  
 A prime example of evolutionary expansion, Gross Domestic Product (GDP) time series  gauge  economic activities   in  
 changing   societal structures,} and 
the  accelerating trend of their  growth 
 probably reflects a  manyfold increase  of the 
  human interactions that drive change. Naively, one could  think of  cultural evolution  as the result of   an optimization process,
 and then expect   the associated GDP growth to   have a 
  decelerating trend. We show how  optimization and expansion  can coexist  by replacing  `wall clock time' $t$ as independent variable 
   with a measure of human interactions intensity $\tau$. The latter was  introduced in a previous work  and is computed 
   using the GDP time series itself.
   
 Our  analysis of  eight centuries of yearly GDP data  from
 three regions of Western Europe, corresponding to present day UK, France and Sweden is  
 carried out in two steps. 
First, a Monte Carlo algorithm is used to fit the GDP data to a piecewise continuous function comprising a
 sequence of exponentials with different exponents.
These arguably correspond to  social and technological stages of societal organization.
 In a second  step, GDP data are plotted vs. $\tau$
 and shown to  display two logarithmic regimes, both decelerating, that are  joined by a power-law cross-over period.
  We connect the end of the first regime and the  beginning of the second
  with the dawn of the Industrial Revolution and the societal impact of new transport, communication and
  production technologies
  that became widely available after  World War I.
  We conclude that wealth evolution in terms of $\tau$ is   a decelerating process with  the hallmarks of record dynamics optimization.
\end{abstract}

\maketitle

\section{Introduction}
{ Optimization and expansion are two modes of  evolution~\cite{RasmussenSibani2019}
   where macroscopic observables change at a decreasing, respectively increasing, rate.
In the first  case, the number of microscopic variables and their interactions
are fixed  while in the second they  change dynamically.  
In this work an example shows how  a variable  change
accounting for the evolving  interactions  reveals 
an optimization process  behind  evolutionary expansion.   }
           
{Human culture is a prime case of evolutionary expansion shaped by  interactions whose intensity has
 changed manifold  due to population growth and  increasing means of communication. Its evolution is mirrored by
 Gross Domestic Product (GDP) pro capita~\cite{Costanza14}, a measure of economic activity and, we surmise,
of  human activity in general.} 

 A culture is defined by its artifacts, from cathedrals to computers, by the technical know-how needed
  to produce them, from masonry to electronic engineering, and by the laws and traditions that express shared values and regulate social behavior, 
  from the Danish fiscal code to the Italian cuisine.
All these things taken together are here dubbed wealth and denoted by the symbol $w$. They are the  products of
preceding economic activity and shape the  framework for its succeeding development.
 
  The long time series~\cite{OWD,Maddison18} that record annual GDP pro capita   in present day dollars  are a  useful proxy for societal wealth history.
The nature of the growth trend as well as the economic  fluctuation around it have attracted considerable attention over the years, see e.g.
~\cite{Kondratiev1925,VDuijn1983,RasmussenMosekildeSterman1985,Sterman1986,MosekildeRasmussen1986,RasmussenMosekildeHolst1989}, and more recently by Ref.~\cite{RosenlystSiboniRasmussen2018} 
and by the present authors~\cite{SibaniRasmussen2020}, who used GDP data to test a simple causal and strongly aggregated model.

The present work has two parts.
In Section~\ref{Staged exponential growth} a detailed GDP data analysis is presented, treating wealth growth as a socio-technical succession process~\cite{MosekildeRasmussen1986}.
The data trend is therefore modeled by a succession of exponentials having different durations and rates.  
Each section is interpreted as a metastable stage of a dynamical process. 
The transitions from one stage to the next are called `quakes'~\cite{Sibani20a} and allegedly express important societal transformations and/or historical events.
 { The   de-trended fluctuations 
  appear as a stationary process characterized by a power spectrum, an autocorrelation function
and an average stage  length of the order of a human lifetime. 
}

 { An ongoing urbanization process, see e.g.~\cite{Ritchie19}
  intensifies the human interactions that 
generate  innovation and wealth~\cite{Bettencourt07}. 
Interaction  intensity and  GDP history are linked 
by a previously introduced  `interaction' variable' $\tau$~\cite{SibaniRasmussen2020} 
 that  in Section~\ref{Wealth evolution vs interaction intensity} replaces
`wall clock' time as independent variable in our data analysis.}

Wealth evolution in terms of $\tau$ has just two different regimes connected by a rapid cross-over,
 both regimes  featuring  logarithmic growth.
We find a strong indication that quakes are a log-Poisson~\cite{Sibani20a} process, which is the hallmark of a
complex  optimization process. 
An underlying evolutionary optimization process~\cite{RasmussenSibani2019} might therefore drive the evolutionary expansion of  economic wealth. 
The final Section~\ref{s:Discussion} contains a summary and a critical discussion of our findings.

\section{Staged exponential growth}
\label{Staged exponential growth}
\subsection{Method}
\label{s:MethodA}
 \begin{figure*}[t!]
	\vspace{-3.2cm}
	\[ \begin{array}{lr}
	\vspace{-5.5cm}
	 \includegraphics[width=0.45\linewidth]{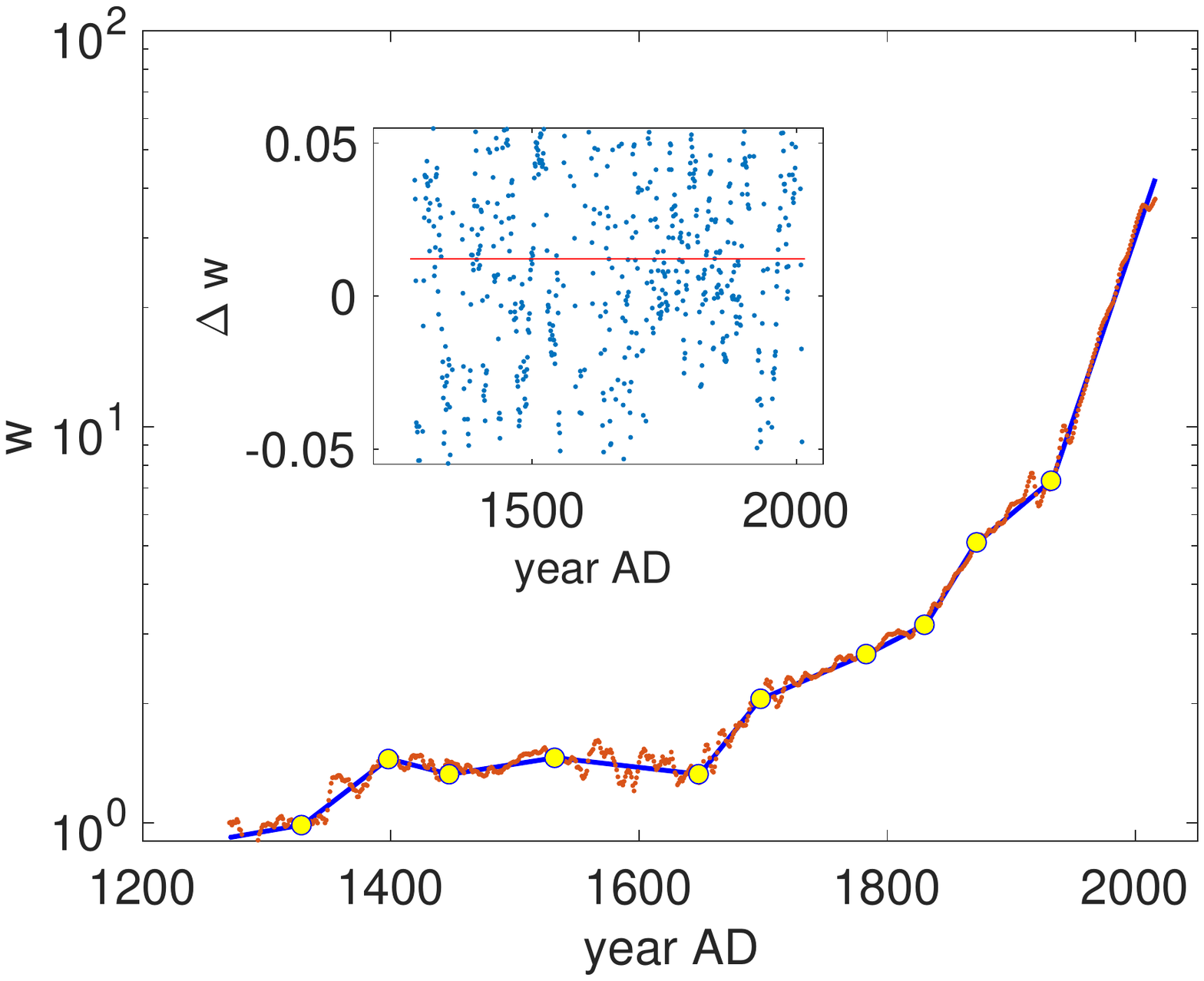}& \includegraphics[width=0.45\linewidth]{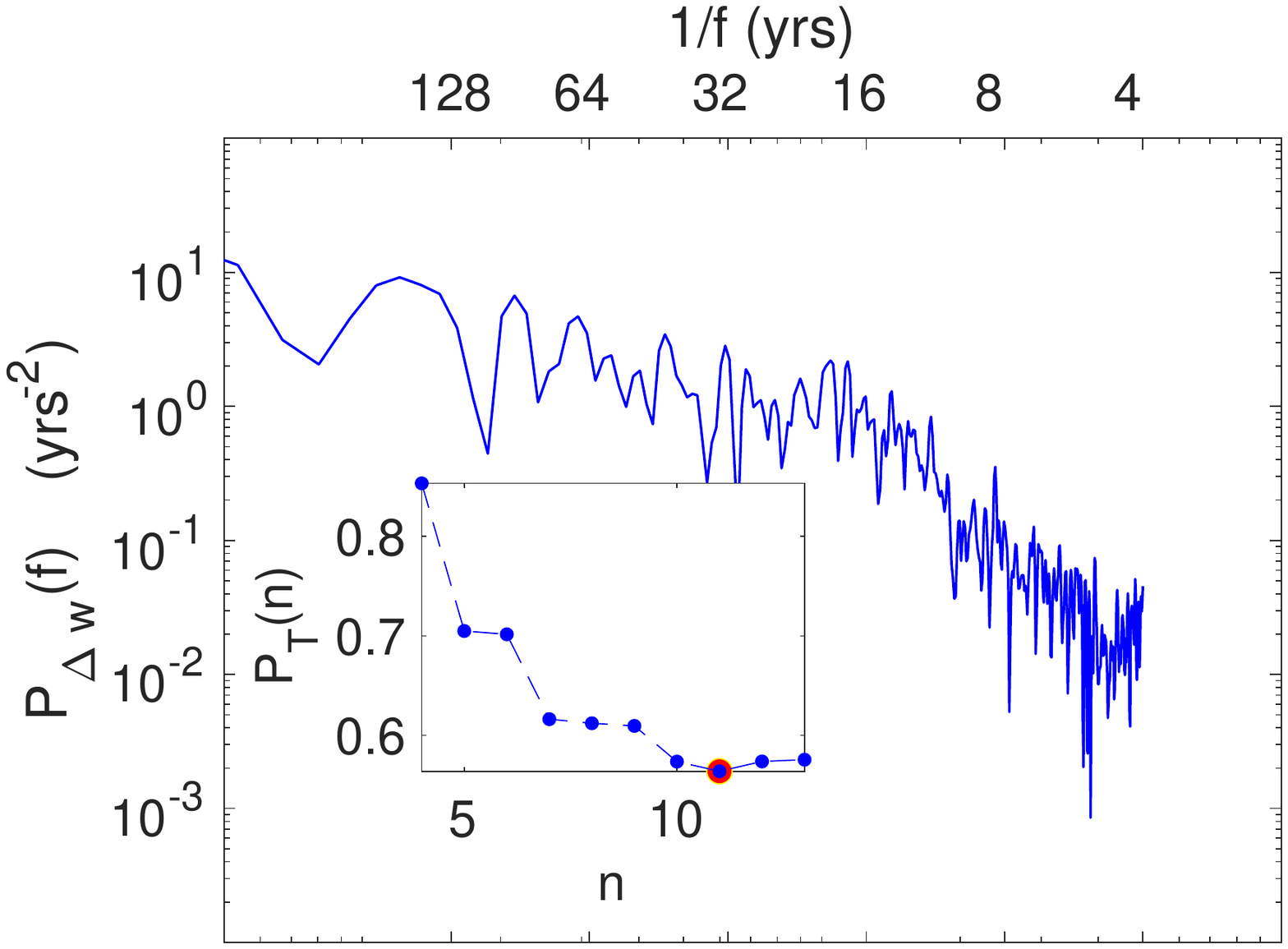}\\
	 \vspace{-5.5cm}
	 \includegraphics[width=0.45\linewidth]{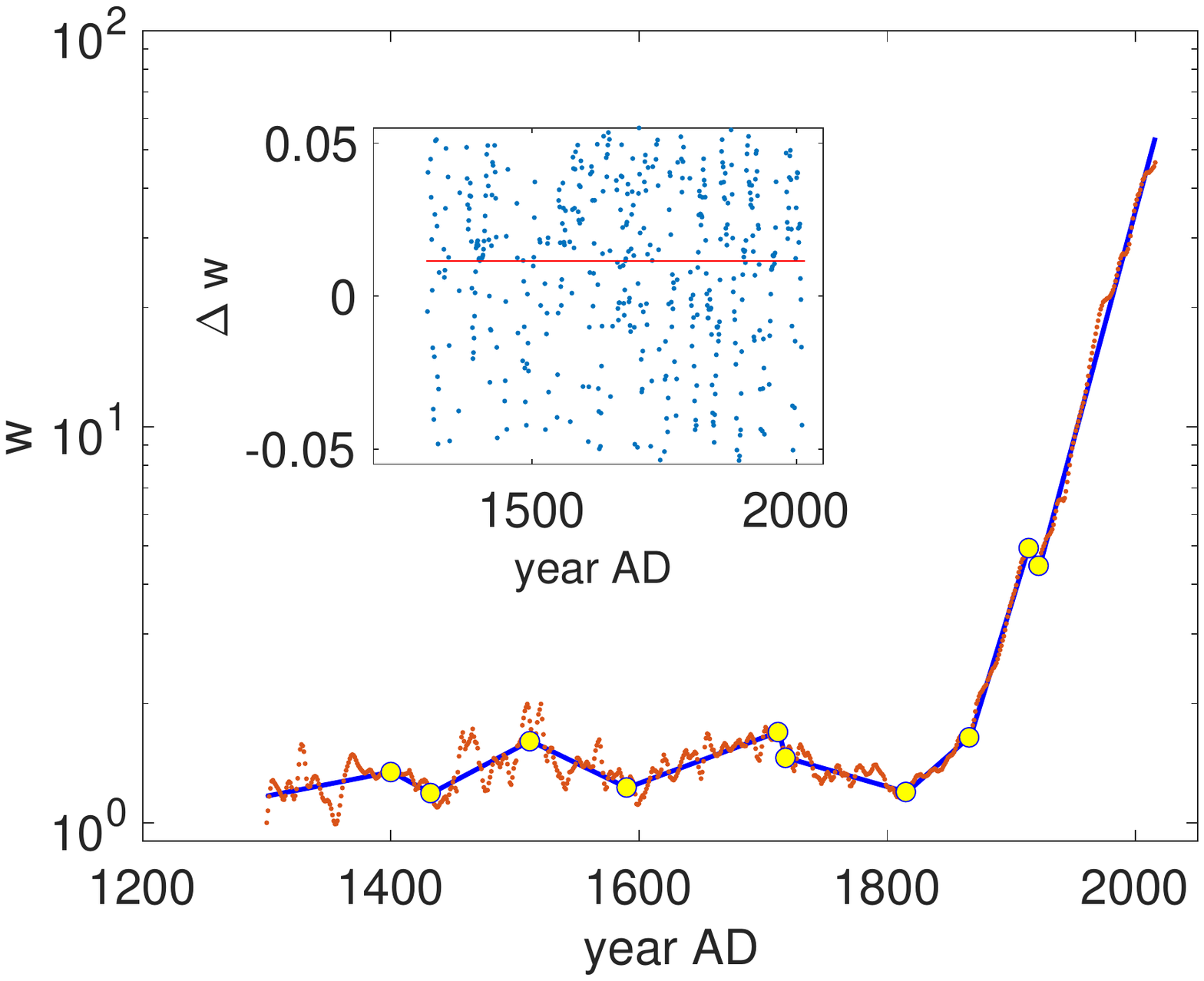}& \includegraphics[width=0.45\linewidth]{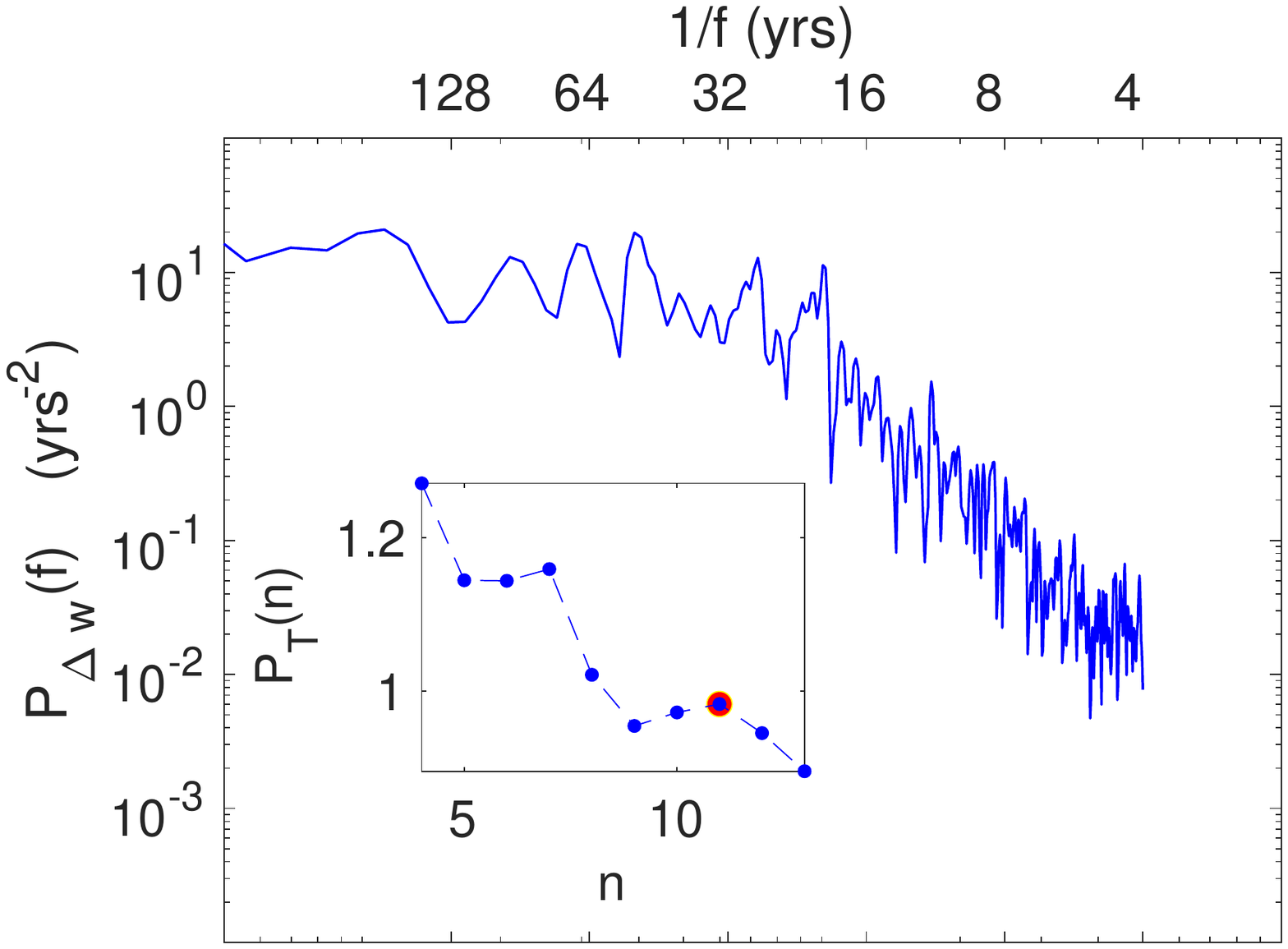}\\
	 \vspace{-5.5cm}
	 \includegraphics[width=0.45\linewidth]{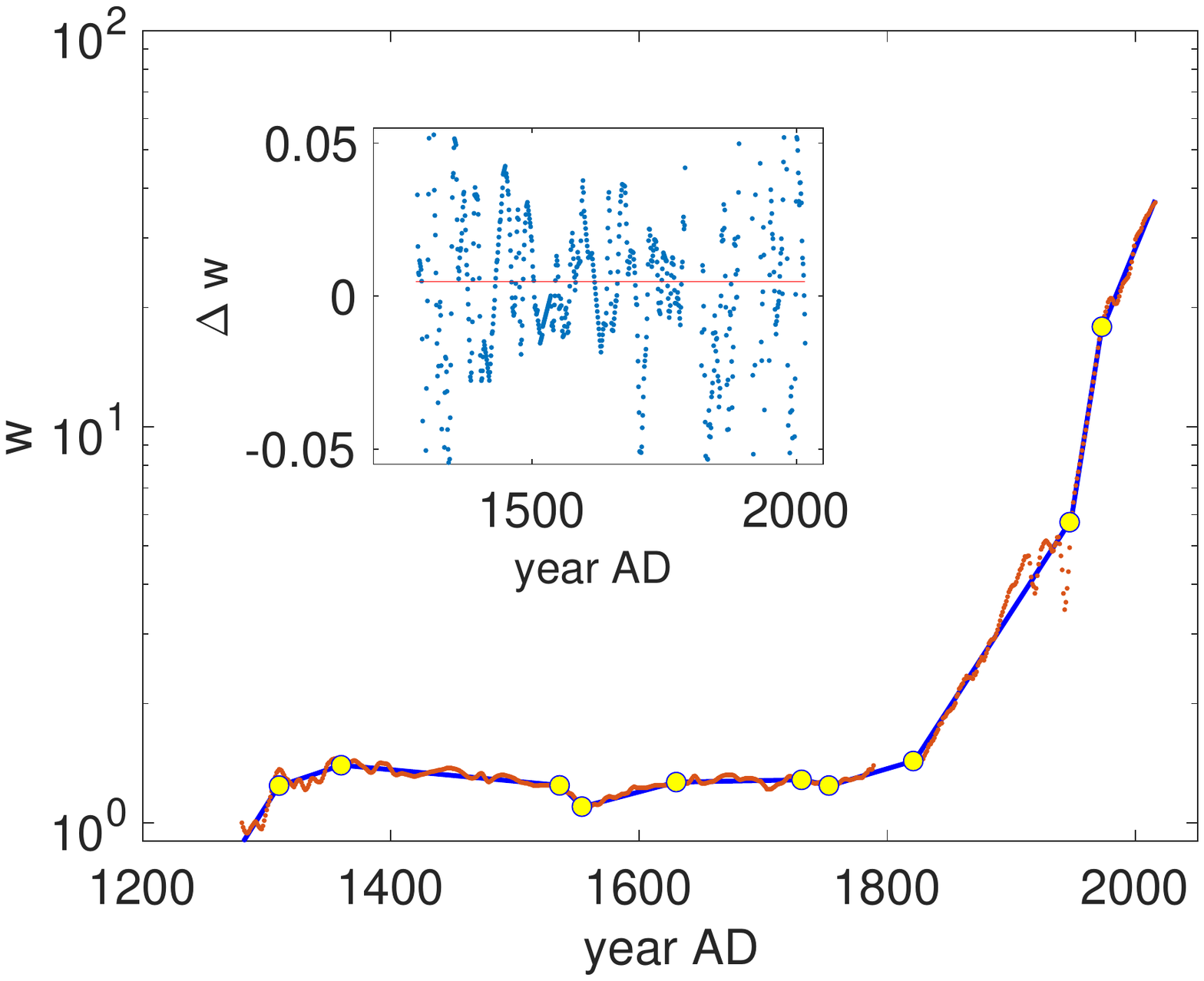}& \includegraphics[width=0.45\linewidth]{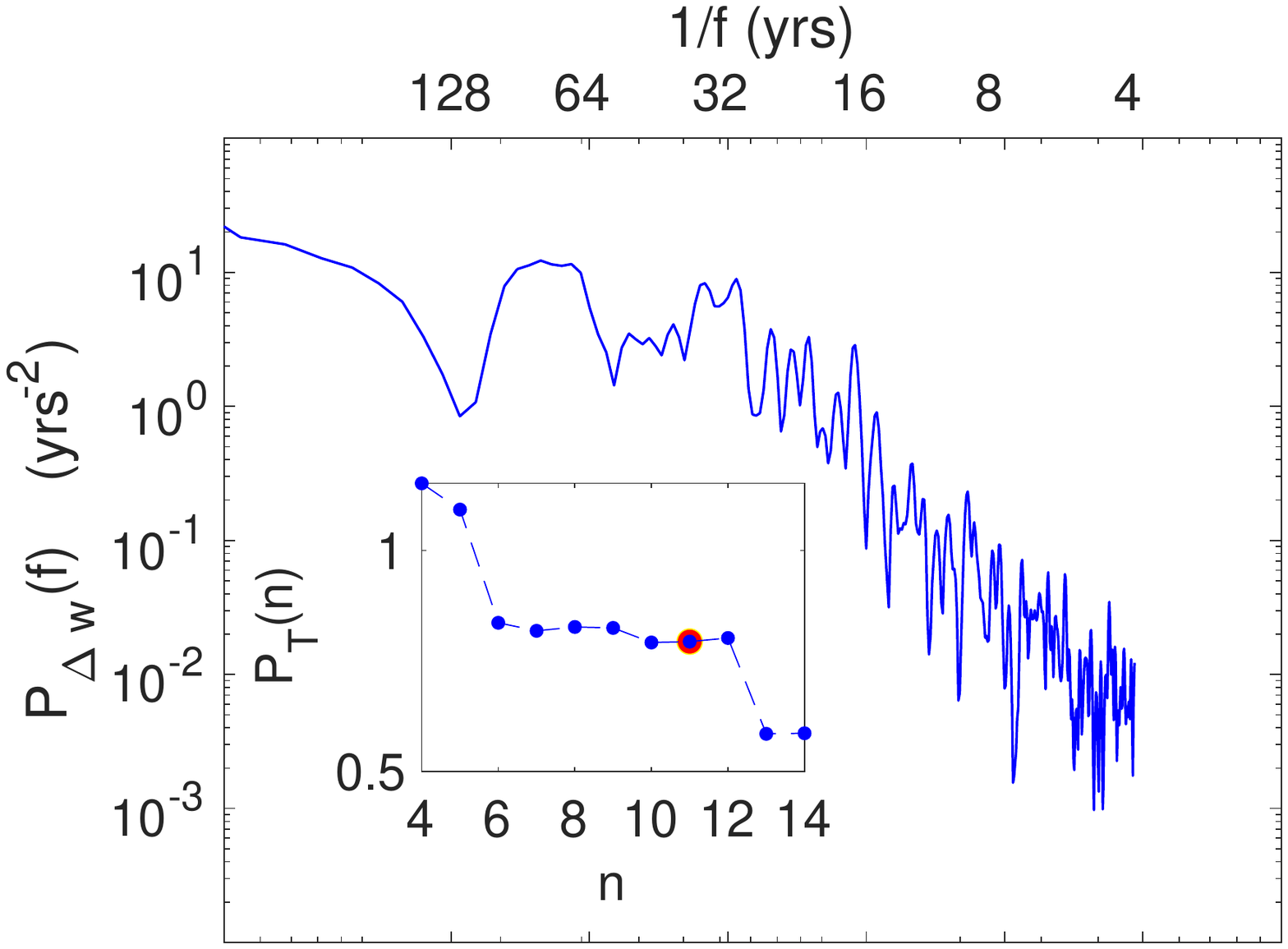}\\
	\end{array}{lr} \]
	\vspace{2cm}
	\caption{Top to bottom: data analysis of the logarithm of GDP pro capita for regions corresponding to modern UK, Sweden and France. 
	Left column: Black lines depict a piecewise continuous  fit, $y(t)$, composed by eleven line segments, each corresponding to an exponential function. 
	The yellow  circles indicate the start of a new segment or stage in the GDP evolution process.
	The red jagged line depicts  the GDP  time series  normalized by its initial value. 
	Insert: the logarithm of the ratio of the GDP time series to its fit is plotted vs. time.
	Right column: Power spectrum as a function of inverse frequency, for $n=11$.
	Insert: The 'goodness of fit'  $P_T(n)$ vs. the number $n$ of stages   depicted as  blue circles. 
	 Each  point is  the square root of the integral of the power spectrum.
	Lower values mean better fit. 
	The $P_T(n)$ value for  $n=11$, the number used in the data analysis presented, is highlighted by a larger blue circle enclosed in a red annulus.
	Data are from~\cite{OWD}.
	}
\label{fig:wealth3C}
\end{figure*}
Time series of GDP pro capita gauge economic activity, but, as mentioned,  they are   treated here as  proxies for `wealth', 
a quantity that, far beyond tradable assets, includes the underpinning social structures, culture and know-how framing human endeavors. 
All time series, called $w(t)$ for wealth, are scaled by their initial value.
As a first step, the logarithm of the series is fitted by a  piecewise continuous linear function comprising $n$ segments, see e.g. Fig.~\ref{fig:wealth3C}.
The resulting trend  $y(t)$  is thus a continuous piecewise smooth function consisting of $n$ exponentials. 
Each of these describes a metastable state, or stage, of the evolving economy.

Let  $q_1$ mark the initial time of the series, and $q_l$, $l=2,3,\ldots$, index the time of the quake connecting stages  $l-1$ and $l$. 
This  quake also marks a discontinuity, or `breakpoint',  in the derivative of $y(t)$.
The duration of the last stage is determined by the end time of the  series, while all other stages have length $p_l=q_{l+1}-q_{l}$. 

 { For given $n$,  the   fitting procedure  described below determines the best  placement of  the $n-1$ breakpoints of $y(t)$
 by minimizing the distance between the log of the trend and the log of the GDP data. 
  We note that the  distance  $P_T(n)$ corresponding to optimal  placement mainly decreases with increasing  $n$. 
  The improvements are initially large and diminish in stepwise fashion as $n$ increases, see right hand column of Fig.~\ref{fig:wealth3C}.
Choosing $n$ large enough would produce a $y(t)$ consisting of many short pieces and enable one to closely follow  the 
details of the data.
  A perfect `fit' can e.g. be obtained for $n$ equal to the length of the time series. }

 { Overly emphasizing fit quality seems  beyond the point as  for stages to qualify as  metastable states
   their  duration  should be, at least in an average sense,
 of the same order as a human life time.
Furthermore, details  reflecting  e.g. business cycles and random events   should not be incorporated in a trend.
    A reasonable choice of $n$  is therefore   the 
 smallest value ensuring  a  good fit and producing stationary de-trended  fluctuations. Finally, we chose to use the same $n$ value for all
 countries.
  Our previous investigations of these and related GDP time series support an $n$ between 9 and 13 as we there found auto correlation functions
   with clear exponential decay times of 50-70 years \cite{SibaniRasmussen2020} as well as peaks in the power spectra at 50-70 years \cite{RasmussenMosekildeHolst1989,RosenlystSiboniRasmussen2018}. 
  Slight variations of our choice, $n=11$,  
 have minor effects on  the integral of the power-spectrum and on  the number of quakes and   do  not qualitatively change our results.
  In particular, that  the stage duration shown in the right hand panel of Fig.~\ref{fig:3fcompare} has no detectable systematic dependence
  on  wall-clock time is a property  unaffected by small variations of $n$.  As a consequence, the logarithmic trend shown in the right hand panel of Fig.~\ref{fig:no_q}
  is  robust to the same variations.
 }
 The above choice of $n$ is  further supported by previous investigations of the long term GDP evolution
  dynamics~\cite{SibaniRasmussen2020,Kondratiev1925,VDuijn1983,RasmussenMosekildeSterman1985,Sterman1986,MosekildeRasmussen1986,RosenlystSiboniRasmussen2018}
  that indicate that the duration of a metastable state, at least in an average sense, is of the  order of magnitude of a human life time.

 For given $n$,  the fitting procedure must determine the placement of  $y(t)$'s $n-1$ breakpoints.
 The $n$ exponential functions forming $y(t)$ have the form $A_l e^{\alpha_l t}$, 
 where the pre-factors, collectively termed $\mathbf A$, are determined by continuity and by the initial condition $A_1=1$. 
 The exponents, collectively denoted by $\mathbf \alpha$, are determined by Monte Carlo (MC) optimization. 
 For fixed $n$, the set of all pairs ${\mathbf A,\alpha}$ defines the configuration space of the MC procedure.
 The best value of $n$ is chosen \emph{a posteriori}, after performing separate optimizations for a range of $n$ values $n=3,4,\ldots,14$. 
 \begin{enumerate}
 \item For a given placement $\bf q$ of the break-points on the time axis, calculate $\bf \alpha$ by minimizing the norm of the logarithmic difference $\ln(y(t)/w(t))$. 
  { The Matlab \emph{fminsearch} function is used for this step. Subsequently,
  the mean distance between consecutive breakpoints is calculated, and to each section of the fit  a small penalty is assigned proportional
 to the difference  between  its length and  the mean distance. This slightly favors equidistant points.}
  The error $E({\bf q,\alpha})$ corresponding to the best fit is saved.               
 \item Optimize the placement of breakpoints  $\bf q$: candidate placements are generated by randomly picking a breakpoint and randomly changing its position by $\pm1$ year.
 The error $E({\bf q,\alpha})$ for the candidate configuration and its difference to the error of the current configuration is calculated $E_i - E_{i-1} = \delta$. 
 The Metropolis acceptance criterion is used, i.e. the move is accepted with probability $P=\min(\exp(-\beta \delta),1)$.
 The inverse temperature $\beta=200$ was used in all simulations.
 \end{enumerate}
For each $n$ the procedure yields a trend $y(t)$, a power spectrum of the de-trended data, and the square root of the integrated power spectrum. 
The latter is a measure of the intensity of the fluctuations.

\subsection{Results}
Our fitting procedure is applied to time series~\cite{OWD} of GDP per capita in a geographical regions corresponding to our modern UK, Sweden and France.
The GDP data scaled to initial value one are shown  as $w(t)$  by red dots in the left hand column of Fig.~\ref{fig:wealth3C}, together   with their respective trends $y(t)$.
The trends are plotted using black lines, and  the quakes are marked by 
 yellow circles. Note the logarithmic ordinate.
 Fluctuations, dubbed $\Delta w$, are shown in the inserts. They are calculated as differences between the logarithms of trends and data and their power
 spectra are shown in the right hand column of Fig.~\ref{fig:wealth3C}.
 The square root of the integral of the power spectra, i.e. the $L^2$ norm of the fluctuations,  are shown  vs. $n$  as inserts of the
 power spectra plots.
 The chosen value $n=11$ is highlighted with a larger red symbol. The choice is good  
  for the UK data shown in the top row, while $n$ = 9 and $n$ = 13 would be better 
   for Sweden and France respectively. 
   {
    The most significant features of the power spectra in Fig. ~\ref{fig:wealth3C} are:  
  Higher power for low frequencies ($1/f > 20$ years) than for high frequencies ($1/f < 20 $ years).
  Low frequency part of the spectra are at a close to constant level, which indicates a lack of  significant correlations. 
  A decreasing slope for high frequencies  indicates the presence of some correlations in that region. 
  When comparing the spectra in Fig. ~\ref{fig:wealth3C} with those  obtained by subtracting a single globally estimated trend from the same time series~\cite{SibaniRasmussen2020}, we see
  that  the staged trend removal process reduces the low frequency 
   power by three orders of magnitude. Furthermore,
  the stage  lengths identify the 'long waves' in the GDP data. }
                      
\begin{figure*}
	\vspace{-3.5cm}
	 \includegraphics[width=0.45\linewidth]{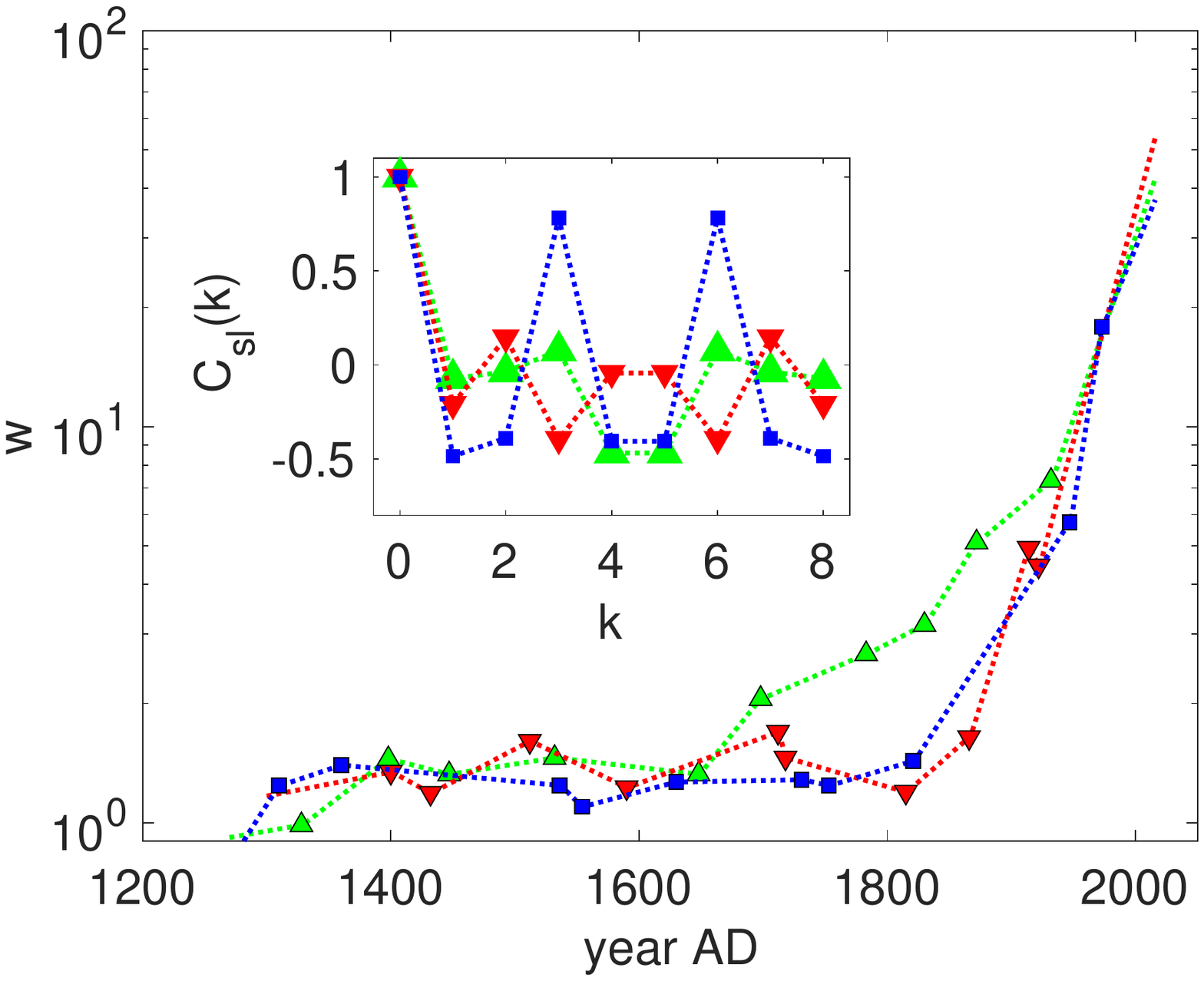}  \includegraphics[width=0.45\linewidth]{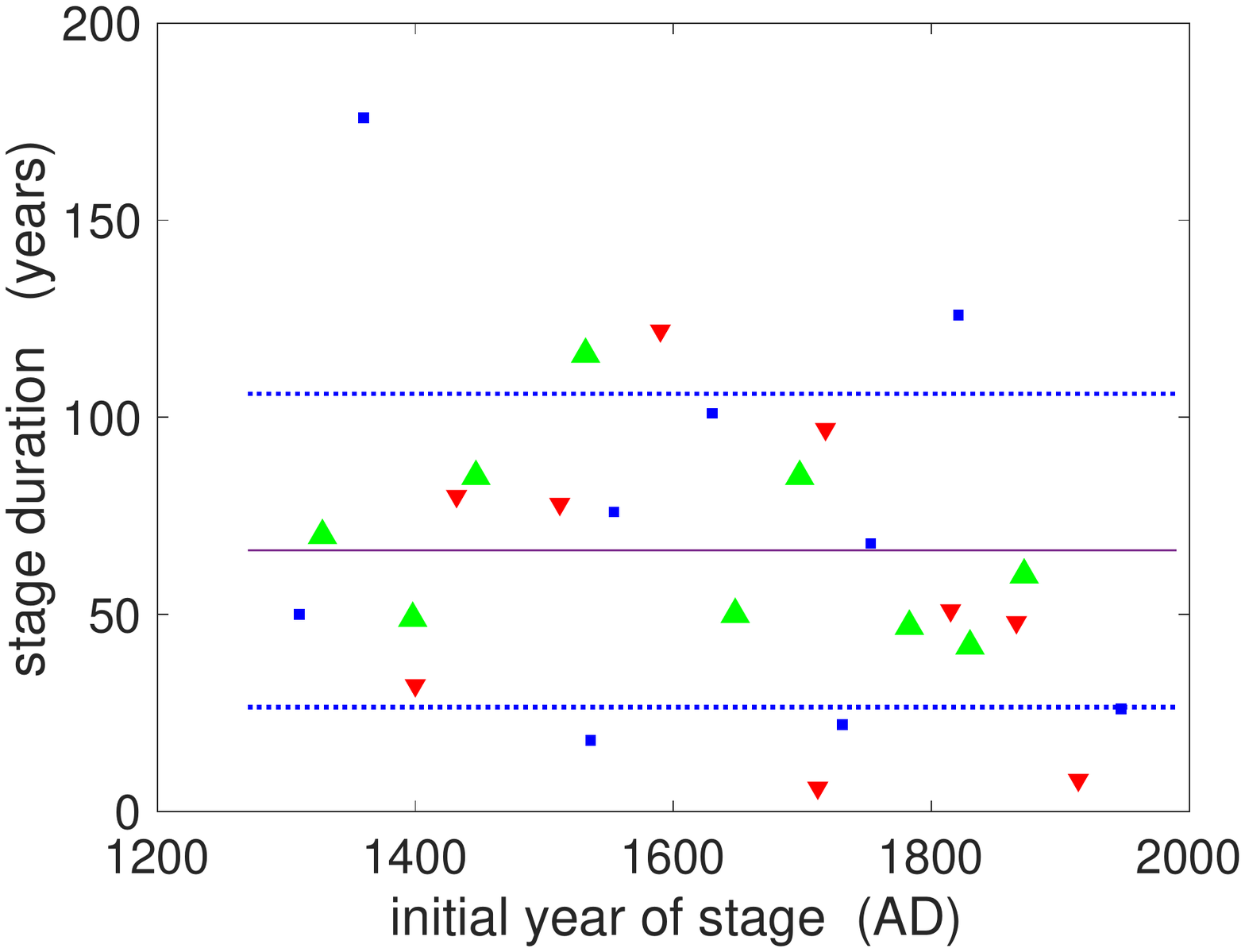}
	\vspace{-3cm}
	\caption{ Left: The GDP fitted trends for UK (green ), Sweden (red ) and France (blue).
	  In the main figure, the  trends are plotted vs time
	with the transition points between different stages  highlighted by up- and down pointing triangles, and squares.
	The insert shows the sample autocorrelation function $C_{sl}$ for the ordered stage length sequence.
	The abscissa $k$ is the lag variable and the lines are only a guide to the eye.\\
	Right: Scatterplot of stage lengths vs.  stage  initial times.
	The full line indicates the mean length  and the two dotted lines the mean $\pm$ one standard deviation. 
	Same symbol and color codes as the left hand side.
	}
\label{fig:3fcompare}
\end{figure*}
\vspace{0cm}

Oscillatory behavior of shorter GDP time series is widely discussed, see e.g.~\cite{Yegorov11,Korotayev10}. 
 Cycles with a period of 7 - 11 years are usually identified as business cycles, those with periods of 15 - 25 years are associated with Simon Kuznets~\cite{Kuznets1930}, and long cycles of 50 years or more with Nikolai Kondratiev~\cite{Kondratiev1925}. 
These oscillations seem to be present in our power spectra although long cycles are virtually gone compared to Figs. 5 
and 6 in ~\cite{SibaniRasmussen2020}, being replaced by different stages, as explained further below.

 Figure \ref{fig:3fcompare}, left hand panel, shows the GDP fitted trends $y(t)$, with different colors and symbols used for the three regions considered.
The corresponding stage lengths form an ordered series, whose sample autocorrelation $C_{sl}$ is plotted
as a function of the lag variable $k$ in the insert of  Fig.~\ref{fig:3fcompare}.All lines are just guides to the eye.
It is clear that  Sweden and France   follow each other closely
and trail the UK for a couple of centuries.
The sample autocorrelation  for the UK data  shows  weak correlations. This is not the case for France, where  several short stages
are the effect of very large GDP fluctuations. Sweden is in this respect in-between the other two countries.

The right hand panel  of Fig.~\ref{fig:3fcompare} is a scatterplot of  stage durations vs. stage  initial  times.
 The full line indicates the average stage
 length  and the two dotted lines the average shifted by one standard deviation.
Note  that  stage lengths  do not  show a systematic change over time
and that  the average stage length, just below 70 years,  is close to a human lifetime. 
This provides an additional time scale that is
connected to regime changes rather than to   oscillatory behavior.

\section{Wealth evolution vs interaction intensity }
\label{Wealth evolution vs interaction intensity}
\subsection{Method}
Cultural evolution is brought about by human interactions that have increased manyfold over time, thanks to population growth, a steady process of urbanization~\cite{Ritchie19} and ever faster and more efficient means of communication~\cite{Schlapfer14}.
 The rapid and substantial change over time of the interaction intensity suggests an alternative to `wall clock time' $t$ as independent variable of the evolution process.
 
 A variable $\tau$ describing the frequency of human interactions was introduced in~\cite{SibaniRasmussen2020}.
There we argued that  a fraction $\gamma w(t)$ of the current wealth is used to improve communication intensity, where $\gamma$ is a positive constant. 
This means $\frac{d\tau }{dt}=\gamma w(t)$ and, by integration,
 \begin{equation}
  \tau(t)= \gamma \int_0^t w(z)dz
  \label{taudef}
  \end{equation}
  
  The rapid fluctuations of the  data  are  smoothened out by the integration, leading to the monotonously increasing form of $\tau(t)$ seen in Fig.~\ref{fig:tauVSt}.
  On sufficiently short time scales, $w$'s trend is nearly constant, and $\tau(t)\propto t$.
  On longer time scales the increasing interaction intensity  leads to $\tau$ growing faster than $t$.
    Importantly,  $\tau$ depends on the cumulated wealth and retains a permanent memory of past events. 
Permanent memory  seems reasonable since  forgetting past achievements, e.g. the printing press, the transistor, or women's rights to vote, requires a near total destruction of human society.
{Finally, varying the value of $\gamma$  rigidly shifts the wealth vs. $\tau$ dependencies shown in Fig.~\ref{fig:no_q}, but does  not affect the 
`wall-clock' time values of the transitions see left hand panel of  Fig.~\ref{fig:no_q} and Table ${\mathbf I}$.
 This was numerically verified using ten different values  of $\gamma$ in the unit interval.
   The value  $\gamma=1$ is used for simplicity.}

 We are interested in the $\tau$ dependence of the wealth $w(\tau)$ and in the quake distribution  on the $\tau$ axis.
 Since both quantities  have a  general form   independent of the value of $\gamma$, the latter is set to one for convenience.
  As shown in Fig.~\ref{fig:tauVSt}, our `interaction' variable $\tau(t)$, obtained by  numerical evaluation of Eq.~\eqref{taudef},
  is close to a linear function of time for the first part of the time series.  
 Once $\tau(t)$ is determined   $w$     can be plotted    vs. $\tau$ without  further assumptions.

\subsection{Results}
\begin{figure}[t!]
	\vspace{-3.2cm}
	 \includegraphics[width=0.95\linewidth]{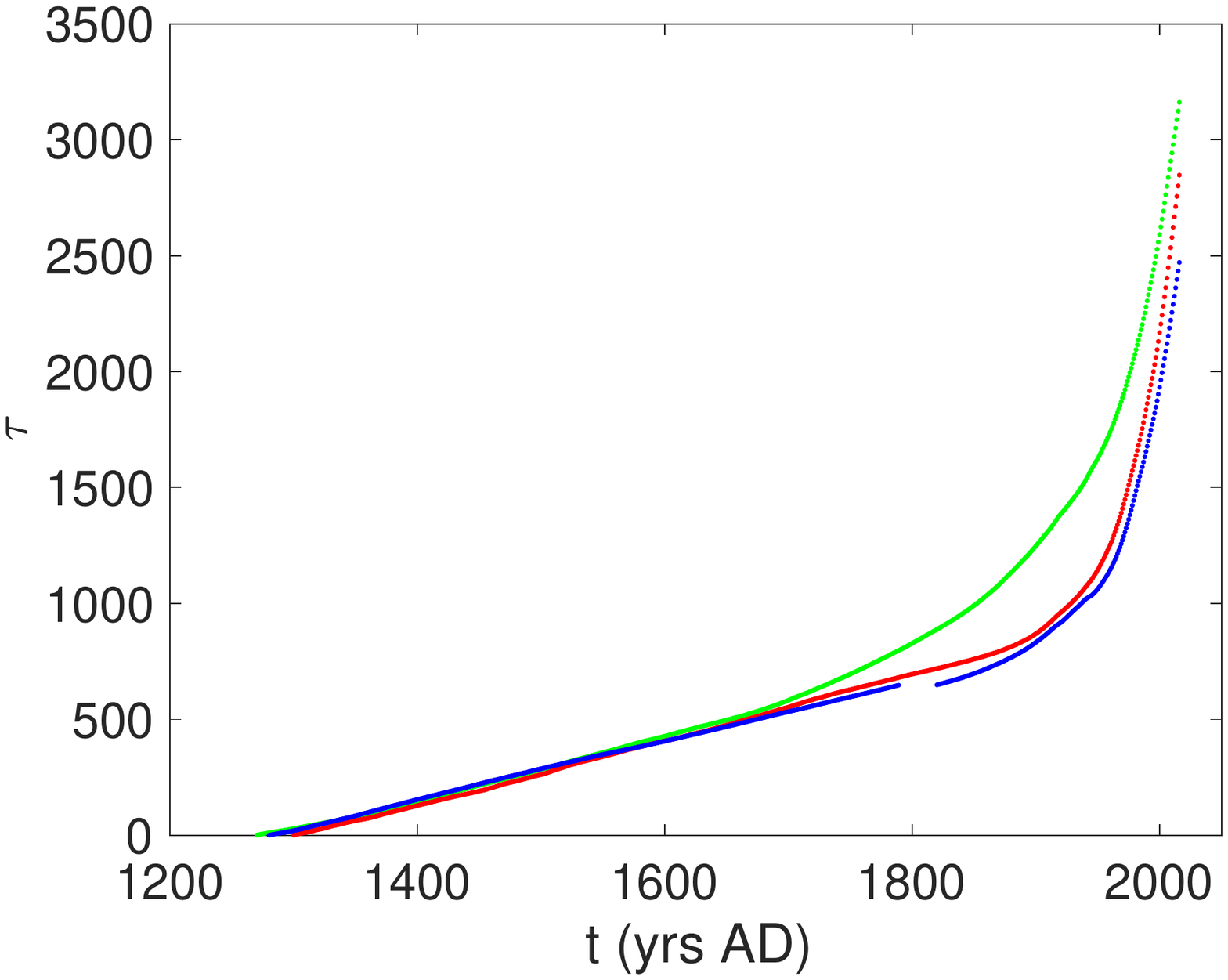}  
	\vspace{-3cm}
	\caption{ The `interaction variable' $\tau$ is calculated by replacing the integral in Eq.(1)
	with a sum. The wealth data used are for the UK (green), Sweden (red) and France (blue).
	{The gap in the blue line is due to missing data.}
	}
\label{fig:tauVSt}
\end{figure}

 The interaction variable  $\tau$
 is  calculated by replacing the  integral
  in  Eq.~\eqref{taudef} with  a sum.
Figure \ref{fig:tauVSt} shows how $\tau$ depends on the wall clock time $t$, with 
 the UK,  Sweden and France  in  green, red and blue, respectively.
In an  initial period, lasting up to $\approx 1600$ AD, $\tau \propto t$, and  all data show nearly  the same linear dependence.
  Thereafter France and Sweden remain close and  trail  the UK in a faster than linear  growth  that lasts until present days.
 
\begin{figure*}[t]
	\vspace{-3.2cm}
	  \includegraphics[width=0.45\linewidth]{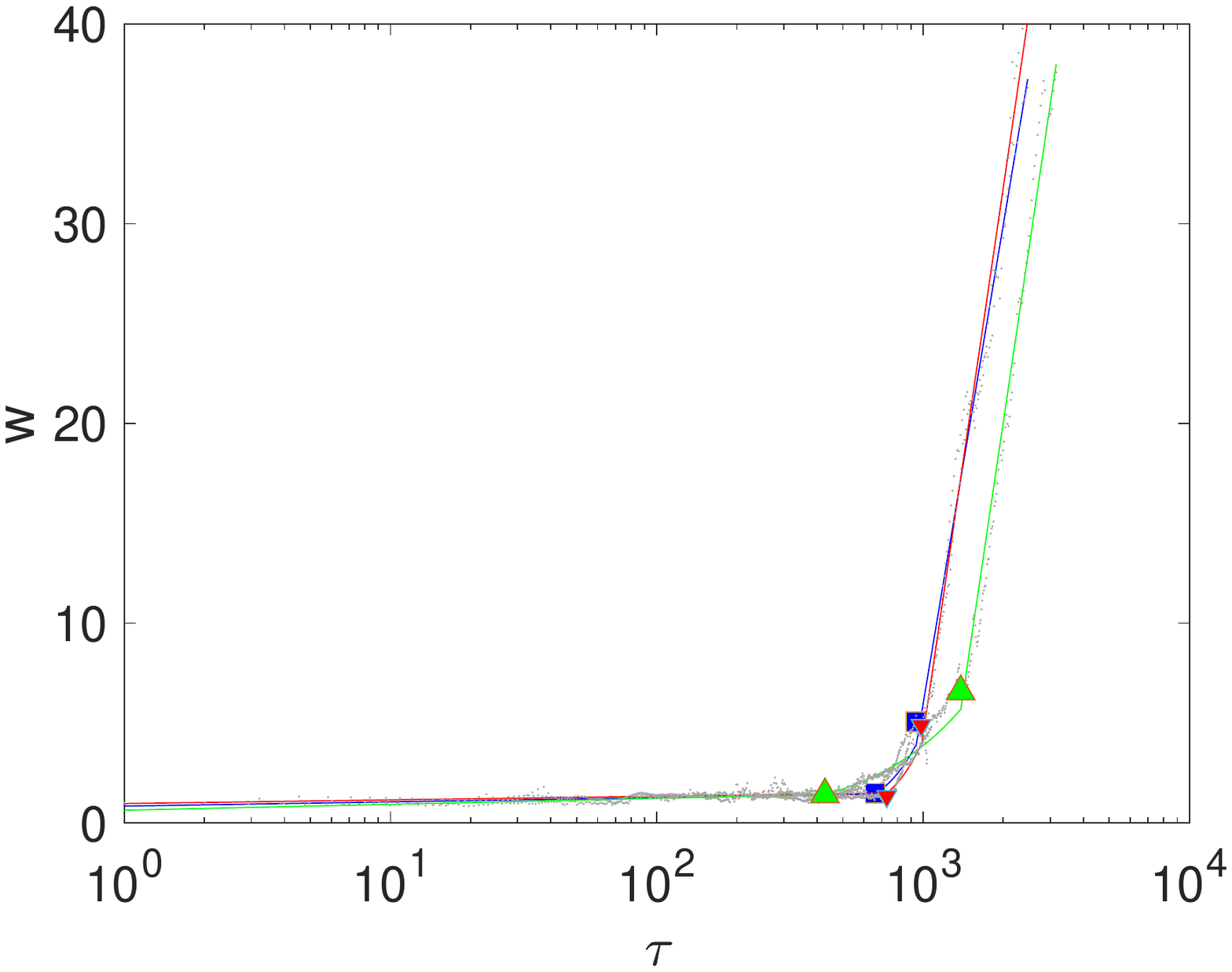}   \includegraphics[width=0.45\linewidth]{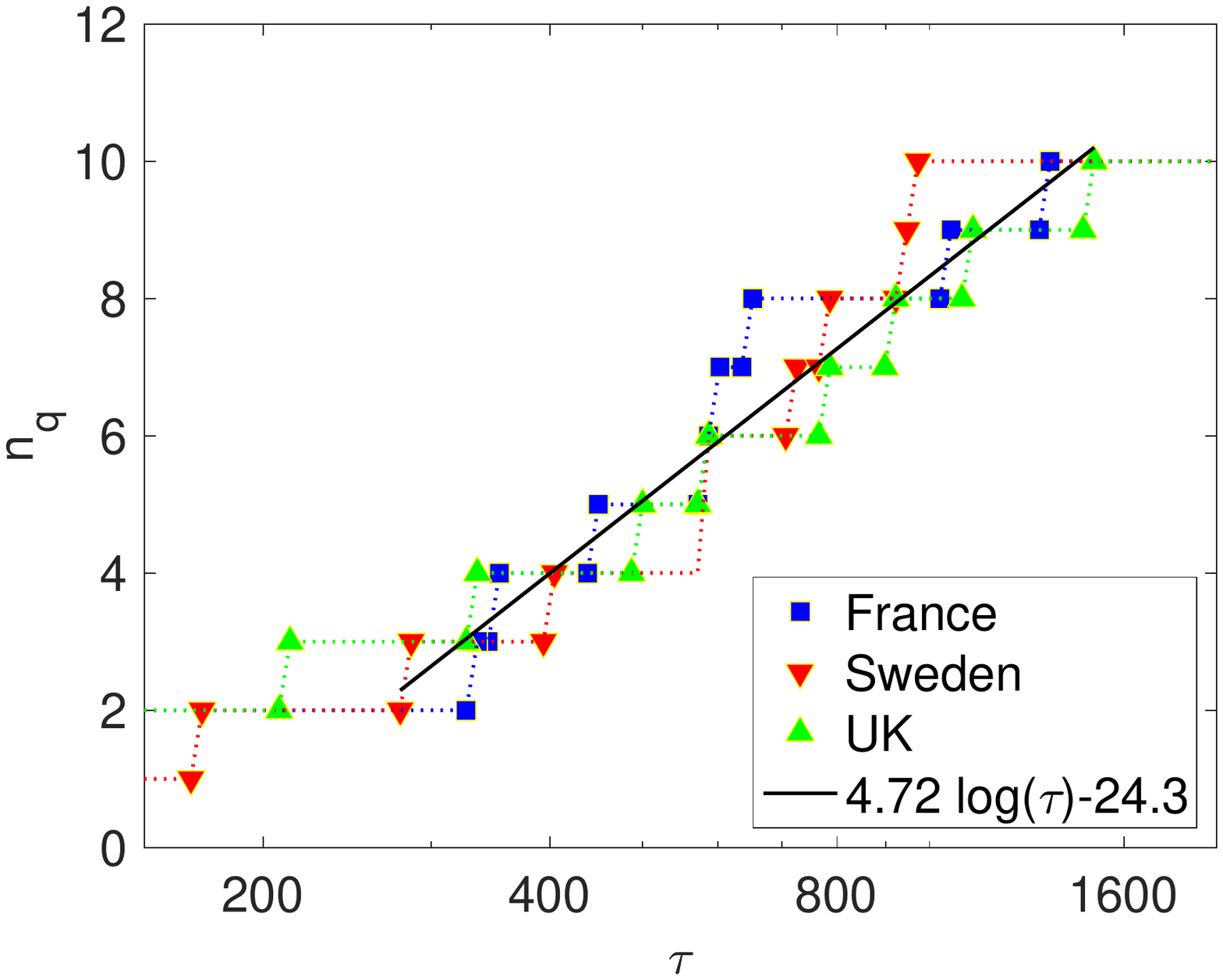} 
	\vspace{-3cm}
	\caption{Left hand panel: UK, Sweden, and France GDP per capita are plotted  vs. $\tau$ with  a logarithmic abscissa
	 using grey dots. Green, red and blue  lines depict  the  fits described in the main text. The symbols, two for each line,
	 mark the transition from the first  logarithmic growth stage to a power law  stage and from the latter to a second stage of logarithmic growth.
	 The transitions occur at $\tau$ values corresponding to $t=1829$ and $1929$ AD for France and  Sweden and to $t=1600$ and
	  $1920$ AD for the UK.
	Right hand panel:  The number of quakes $n_q$ occurring prior to interaction time $\tau$  is plotted vs. $\tau$ using a dotted
	line and a logarithmic abscissa for the  UK, Sweden and France.
	Symbols are placed at the beginning and end of each of $n_q$'s  plateau.
	Note that some of the symbols overlap. 
	The black  line depicts the linear function specified in the legend. 
	}
\label{fig:no_q}
\end{figure*}
 The  color and symbol codes used  in Fig.~\ref{fig:no_q} are specified in the legend of the right hand panel.
 In the left hand panel,   the wealth data  shown in Fig.~\ref{fig:3fcompare} are re-plotted vs. $\ln \tau$, together with 
 their respective fits. The raw data are depicted by grey dots, and the fits by  green, red and blue lines for the UK, Sweden and France, respectively.
Two  regions of logarithmic growth are intercalated by a shorter period with power-law growth.
For each region, the fitting function has the form:
$y=a_1\ln \tau +a_2$  for $\tau< \tau(t_1)$, $y=c  \tau^{a_3}$ for $ \tau(t_1)<\tau<\tau(t_2)$
and $y=a_4\ln \tau + d$ for $\tau>\tau(t_2)$,  The  constants  $c$ and $d$ are determined by the continuity of the fitting  function.
All parameters are given in the table below:
{
\begin{center} 
\begin{table}
\label{tab:parameters}
\[ 
\begin{array}{|| l|r|r|r|r|c|c||} 
\hline
\mbox{} & a_1 & a_2 & a_3& a_4 &t_1 {\rm  (AD)}& t_2 {\rm (AD)} \\ \hline \hline
 \text{UK}&  0.13 &   0.63 &   1.19&  39.2& 1600  & 1920\\
\text{Sweden}  &  0.079   & 0.97 &   3.29&  39.5& 1829 & 1925  \\
\text{France}&0.092  &  0.84&    2.85&   34.6&1829 & 1925  \\ 
\hline \hline
\end{array} 
\]\caption{Parameters of the fit $y=a_1 \ln \tau +a_2$ for $t < t_1$, $y=c \tau^{a_3}$ for
$t_1<t<t_2$ and $y=a_4 \ln \tau +d$ for $t>t_2$. The parameters $c$ and $d$ are determined  by continuity. }
\end{table}
\end{center}
}
   
  Since  wealth vs. $\tau$ grows at a decelerating pace, 
 except in the cross over region connecting the slow to the fast logarithmic phase,
 it behaves as
  an evolutionary optimization process~\cite{RasmussenSibani2019} in most  of its range. 
  
To characterize statistically the quakes identified in the time series analysis, $1000$ logarithmically equidistant points
  are placed on the  $\tau$ axis. For each country, $n_q(\tau)$ is the number of quakes occurring prior to $\tau$.
The function comprises a number of plateaus connected by unit `jumps'. 

The left  hand panel of Fig.~\ref{fig:no_q} depicts   wealth $w$  vs. $\tau$ using grey dots for all raw data and
 green, red and blue lines  for the  fitted trends of the UK, Sweden and France.
 The two turning points of the trend lines are  highlighted by symbols of the same  color.

Note that the symbols for Sweden and France  overlap.

In the right hand panel of the same figure, $n_q(\tau)$, the number of quakes occurring prior to $\tau$,
is plotted using a logarithmic abscissa and green, red and blue dotted lines for the UK, Sweden and France, respectively.
The symbols specified in the legend  highlight the endpoints of the plateaus of the staircase shaped
form of $n_q(\tau)$. The logarithmic fit plotted as a black line and described in the legend,
is based on all the endpoints falling 
in the region $\tau >240$.
That  the $n_q(\tau)$ trend  grows at a constant logarithmic rate, except at the very beginning of the quaking
process, is strongly reminiscent of a Record Dynamics~\cite{Sibani20a} description of complex
optimization.

\section{Discussion and outlook}
\label{s:Discussion}
Eight centuries of GDP time series for the UK, Sweden and France are first analyzed as  staged processes~\cite{MosekildeRasmussen1986,RasmussenMosekildeHolst1989,RosenlystSiboniRasmussen2018,SibaniRasmussen2020} 
and then described in terms of the `interaction' variable $\tau$  introduced in~\cite{SibaniRasmussen2020}.
 Transitions between one stage and the next
 are salient events termed `quakes'.
 Using Monte Carlo techniques, we first identify  the best placement  on the time axis  of $n=11$ quakes 
  and then argue that GDP evolution  is  a decelerating process
   when described in terms of $\tau$. 
 The last step  requires simple numerical operations on  raw data
 and is robust  to small variations of $n$. 
 
 {Eleven stages of exponential growth with mainly positive rates give a good description of the GDP series, with 
  quake positions  that differ across the three countries, see the left hand  side of Figs. ~\ref{fig:wealth3C} and ~\ref{fig:3fcompare}.
  For our choice of $n$, the mean stage duration  lies  slightly below $70$ years and
  the de-trended fluctuations have near zero mean.
  A brief comparison with a  deterministic trend model~\cite{SibaniRasmussen2020} 
 shows that our analysis based on an eleven stage trend qualitatively concurs with the corresponding  results
 of that model.}
  {Having  only three free parameters,  the latter  produces de-trended data with larger variations than in the present case. 
Hence, compared to the present case, the power spectra of  the de-trended time series have  three orders of magnitude higher power for  low frequencies
with strong peaks for $1/f$ between $50$ and $70$ years.
Further,  a clear  $f^{-2}$ background signal is present
that is attributed to the exponential decay of the fluctuation  autocorrelation function with 
decay times   between $50$ and $70$ years.}

{The power spectra in the present analysis  have
higher power for low frequencies ($1/f > 20$ years) than for high
 frequencies ($1/f < 20 $  years) and
 a close to constant low frequency  trend, see  Fig. ~\ref{fig:wealth3C}.
This indicates  a lack of  significant correlations between  low frequency
  de-trended fluctuations.
 A distinct $f^{-2}$ background is missing  and the mean duration of the identified stages lies between
 $60$ and $80$ years for $n$ between 9 and 13 and slightly below $70$ years for the $n$ = 11 case. 
 Furthermore, the  autocorrelation function of the stage duration  is nearly structureless,
  as seen the insert of Fig.~\ref{fig:3fcompare}. 
  The three observations combined suggest that stages qualitatively correspond to the `long waves' in the GDP data appearing as strong low-frequency peaks in
   the power-spectra of Ref.~\cite{SibaniRasmussen2020}. Interestingly, this means that our current analysis 
   also suggests that `long waves', or more correctly, long term rythmic growth patterns, also are detectable in pre-industrial times, 
   while  they originally were thought as being a feature of the industrial economy, see Ref.~\cite{Kondratiev1925}. 
   A decreasing trend for higher frequencies only spans less than a decade and is likely insignificant.}
 
 { A  close correspondence between   quakes and identifiable historical event cannot be clearly  identified.
For example,  the plague outbreak that reached Messina in 1348 AD~\cite{Frith2012}
 falls between two quakes 1328 and 1398 AD in the UK data. After the last event, the UK GDP enters a period of modest
 and even negative growth that ends in 1648 AD, a century earlier than the conventional 
  beginning of the Industrial Revolution~\cite{Wiki_Industrial_Revolution}.}

  In terms of the  interaction variable $\tau$, all wealth trends  have  two periods of logarithmic growth, the first at a lesser rate than the second. 
In the  cross over between the two $w$ is well fitted by a power-law with exponent larger than one. See Table~$\mathbf{I}$.
{ In the UK,  the   transition  start- and endpoints fall more than two centuries earlier  and more than  five years earlier, respectively,
than   in Sweden and France.  That France and  Sweden trail the 
UK's   longer and more gradual transition broadly concurs 
with the fact that the  Industrial  Revolution commenced in the UK.
As mentioned,  the UK transition starts more than 
one century before the `official'   beginning of the Industrial Revolution~\cite{Wiki_Industrial_Revolution}. 
Note that the  UK experienced a significant GDP growth in the 16'th and 17'th centuries
 due to the colonial  activities and trade.
All three  transition  endpoints  fall shortly after World War I. 

   We note that  the fluctuations around the trend line  are  small and  far apart, and not always  temporally close to memorable  historical events. 
   Thus, these  turning  points could likely  be a (time-delayed) consequence of innovations that  \emph{i)} changed  human communication and interaction
   and  \emph{ii)} created the technical basis for the intensive farming needed to support growing urban populations. These mainly
   occurred  during the economic laissez-faire period ending with
     WWI~\cite{Berend2012}. To  the first category belong 
     the electrification of cities, including  urban tram and subway
    systems, the building of  railway and  telephone line networks and, last but not least, the invention of the radio~\cite{Marconi1896}. }
    To the second belong  the production of ammonia on an industrial scale, achieved  in 1913  by the  Haber-Bosch process~\cite{Haber-Bosch1913} and the 
    use of tractors. The first gasoline powered tractors were built in Illinois, by John Charter  in 1889, and 
    the Fordson, a  popular mass-produced tractor was introduced in 1917~\cite{tractor}.

{ Since   the cumulated number of quakes occurring prior to a given $\tau$, see the right hand panel  of Fig.~\ref{fig:no_q}, has a clear logarithmic trend,
GDP evolution expressed in terms of $\tau$  is reminiscent of  the decelerating optimization processes described in 
  Record Dynamics (RD)~\cite{Sibani20a}.  
 In essence, RD links a decreasing rate of change of macroscopic variables to a hierarchy of dynamical barriers
 separating increasingly robust metastable states. 
Once the effect of the ever increasing interaction intensity is removed by replacing $t$ with $\tau$, the structures
 regulating human behavior become indeed progressively harder to change. } For example, 
 modifying existing scientific  knowledge requires progressively more advanced instruments and manpower, 
and the growing body of rules which regulate human interactions, e.g.  legislation, must be  tended by an increasing bureaucracy.
 
 {The  importance  of hierarchies in complex  dynamics  has long been appreciated~\cite{Simon62},
 see \cite{Jiang2022} for a recent application to hierarchical organization of urban space.
 In our case,  the barrier hierarchy  behind evolutionary expansion
 could be  associated  with  how growing social structures 
develop their interactions. 
In conclusion, GDP growth, a manifestation of cultural expansion, is an accelerating process as function of `wall clock' time $t$
but appears as  a sequence of two  decelerating optimization processes each underpinned by a search within a hierarchy,
 when  studied as function of the `interaction' variable $\tau$. 
Open problems are the driving force behind the searches, i.e. the target of the optimization process, 
and the structure of the corresponding dynamical 
hierarchies.  
 }

\bibliographystyle{apa-good}

\end{document}